\documentclass[
twocolumn,
superscriptaddress,
showpacs,preprintnumbers,
amsmath,amssymb,
aps,
prb,
floatfix,
10pt,
]{revtex4-1}

\usepackage{graphicx}
\usepackage{amsfonts,amsmath,amssymb}
\usepackage{url}
\usepackage[utf8]{inputenc}
\usepackage{fancyref}
\usepackage{hyperref}
\hypersetup{colorlinks=false,pdfborder={0 0 0},}

\newcommand{\ket}[1]{|#1 \rangle}
\newcommand{\bra}[1]{\langle #1|}
\newcommand{\bracket}[1]{\langle #1 \rangle}
\newcommand{\siesta}{\textsc{siesta}}
\newcommand{\nemo}{\textsc{nemo\oldstylenums{3}d}}

\begin{document}

\title{Electronic transport in Si:P $\delta$-doped wires}

\author{J.~S.~Smith}
\email{jacksonsmith@tcqp.science}
\affiliation{Chemical and Quantum Physics, School of Applied Sciences, RMIT University, Melbourne VIC 3001, Australia}

\author{D.~W.~Drumm}
\affiliation{Chemical and Quantum Physics, School of Applied Sciences, RMIT University, Melbourne VIC 3001, Australia}
\affiliation{Australian Research Council Centre of Excellence for Nanoscale BioPhotonics, School of Applied Sciences, RMIT University, Melbourne, VIC 3001, Australia}

\author{A.~Budi}
\affiliation{NanoGeoScience, Nano-Science Center, Department of Chemistry, University of Copenhagen, Universitetsparken 5, 2100 K{\o}benhavn {\O}, Denmark}

\author{J.~A.~Vaitkus}
\affiliation{Chemical and Quantum Physics, School of Applied Sciences, RMIT University, Melbourne VIC 3001, Australia}

\author{J.~H.~Cole}
\email{jared.cole@rmit.edu.au}
\affiliation{Chemical and Quantum Physics, School of Applied Sciences, RMIT University, Melbourne VIC 3001, Australia}

\author{S.~P.~Russo}
\email{salvy.russo@rmit.edu.au}
\affiliation{Chemical and Quantum Physics, School of Applied Sciences, RMIT University, Melbourne VIC 3001, Australia}


\begin{abstract}
Despite the importance of Si:P delta-doped wires for modern nanoelectronics, there are currently no computational models of electron transport in these devices. In this paper we present a nonequilibrium Green's function model for electronic transport in a delta-doped wire, which is described by a tight-binding Hamiltonian matrix within a single-band effective-mass approximation. We use this transport model to calculate the current-voltage characteristics of a number of delta-doped wires, achieving good agreement with experiment. To motivate our transport model we have performed density-functional calculations for a variety of delta-doped wires, each with different donor configurations. These calculations also allow us to accurately define the electronic extent of a delta-doped wire, which we find to be at least 4.6~nm.
\end{abstract}

\maketitle

\section{\label{sec:intro}Introduction}

Phosphorus doping in silicon with atomic precision has led to a variety of new nanostructures on the silicon platform~\cite{Zwanenburg2013a}. Foremost among these are metallic wires that are ``one atom tall and four atoms wide''~\cite{Weber2012a}. These wires are made using a $\delta$-doping technique that combines scanning-probe lithography with molecular-beam epitaxy~\cite{Tucker1998a,OBrien2001a,Shen2002a,Simmons2005a}. This technique achieves both high-density carrier concentrations and excellent two-dimensional (2D) confinement of phosphorus atoms in silicon~\cite{McKibbin2010a,McKibbin2014a}. For example, 2D doping densities of one in four inside a (001) silicon monolayer (\textit {i.e.} 0.25~ML) have previously been reported~\cite{Wilson2006a}. These high donor densities result in spatially confined electron transport when an in-plane voltage bias is applied to the nanostructure~\cite{Ruess2008a,Weber2014a}. Therefore, these systems have similar quantum mechanical properties to undoped silicon nanowires which are confined structurally~\cite{Ko2001a,Svizhenko2007a,Ryu2015a}. The electronic properties of Si:P systems have applications for quantum computing and quantum communication technologies~\cite{Kane1998a,Cole2005a,Hollenberg2006a}. In this paper, we examine some of these properties for a phosphorus in silicon (Si:P) $\delta$-doped wire~\cite{Ruess2007a,Ruess2007b,Ruess2008a,Weber2012a,McKibbin2013a,Weber2014a} using density-functional theory (DFT), effective-mass theory (EMT), the nonequilibrium Green's function (NEGF) formalism, and tight-binding (TB) theory.

The long spin-coherence times and large Bohr radius of the phosphorus donor electron in silicon make this material an interesting candidate for spin-based devices and other nanoscale electronics~\cite{Kohn1955a,Faulkner1969a,Tyryshkin2012a,Steger2012a}. Si:P $\delta$-doped wires might be used as the interconnects between stationary and flying qubits~\cite{DiVincenzo2000a,Ruess2007a}, low-resistivity source-drain contacts for nanoelectronics~\cite{Clarke2008a,Weber2012a}, or one-dimensional (1D) spin chains for confined magnon transport~\cite{Balachandran2008a,Makin2012a,Ahmed2015a}. The modern $\delta$-doped wire is a quasi-1D row of phosphorus atoms oriented in the [110] crystallographic direction, with a width of 1.54~nm in the lateral [1$\bar{1}$0] direction which is equivalent to two dimer rows (2DR) on the reconstructed (001) silicon surface~\cite{Weber2012a,Weber2014a}. In addition, because the placement of phosphorus atoms on the (001) plane is indeterministic~\cite{Wilson2006a,Schofield2006a}, a variety of donor configurations are possible within a realistic wire.

Computational models of Si:P $\delta$-doped wires are limited to our density-functional model (Ref.~\onlinecite{Drumm2013a}) and the empirical TB model of Ref.~\onlinecite{Ryu2013a}, where the \nemo~package is used to describe the equilibrium electronic properties of a Si:P $\delta$-doped wire~\cite{Weber2012a}. There are currently no computational models of electron transport in a $\delta$-doped wire. However, electronic transport in an Si:P $\delta$-doped layer has previously been investigated using a semi-classical model~\cite{Hwang2013a}.

In this paper we calculate the electronic properties of a Si:P $\delta$-doped wire for a variety of donor configurations. We investigate the effects of donor disorder and accurately define the electronic extent of a $\delta$-doped wire. The current-voltage (I-V) characteristics of two $\delta$-doped wires have recently been reported experimentally~\cite{Weber2014a}. Therefore, we expand on our density-functional model by using it to develop the first computational model for electronic transport in a $\delta$-doped wire. We calculate the I-V characteristics for the two $\delta$-doped wires which have been reported experimentally and investigate the change in these characteristics for wires with larger in-plane widths. This computational model has wide applicability because it scales easily to large system sizes, which are needed to simulate realistic devices.

\section{\label{sec:dft}Equilibrium electronic properties of $\delta$-doped wires}

In this section we present the results of our density-functional calculations for the equilibrium electronic properties of a variety of $\delta$-doped wires. The equilibrium properties are those for which the potential difference between the source and drain contacts is equal to zero. Of particular interest are those properties relating to the spatial confinement or electronic extent of the donor electrons perpendicular to the axis of the wires. We make an important distinction between two closely related concepts: the \textit{electronic width} and the \textit{electronic extent} of the wires. The electronic width is defined as a measure of spread for the probability density of the donor electrons perpendicular to the wire axis (for example, the full-width half maximum of this probability distribution). This width is sometimes referred to as the ``effective electronic diameter'' of the wires~\cite{Weber2012a}. The electronic extent of the wires is defined as the minimum distance of lateral separation at which two wires do not affect each other's equilibrium electronic properties. Therefore, the electronic extent is the minimum distance at which two wires must be separated so they behave exactly as they do in isolation. By contrast, the electronic width is the region within which it is most likely to find the donor electrons.

\subsection{\label{sec:methods-dft}Density-functional theory}

\begin{figure}[b!]
    \centering
	\includegraphics[]{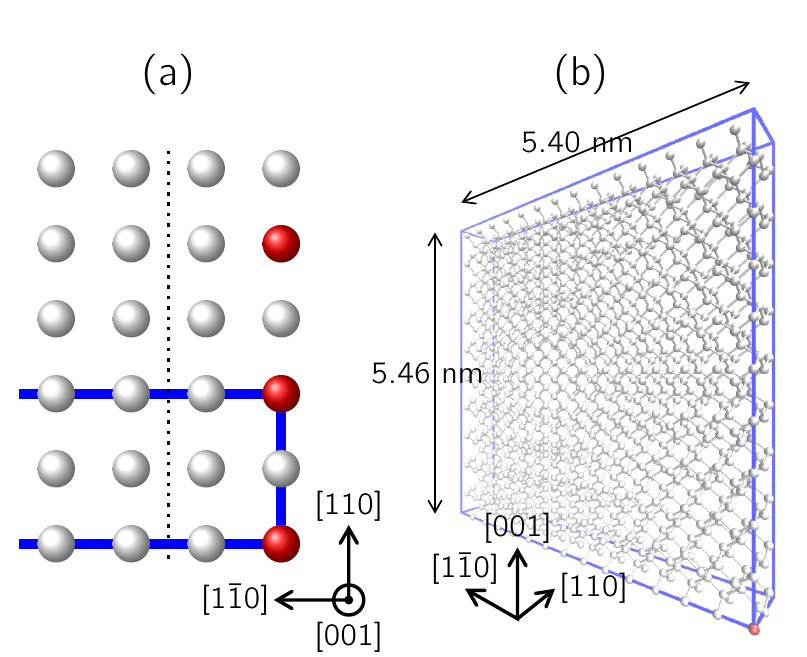}
	\caption{(color online) (a) A 2D schematic of the single-row wire showing only part of the donor plane, and (b) a 3D schematic of an orthorhombic supercell for the single-row wire, with silicon atoms (white spheres), phosphorus atoms (red spheres), the periodic boundaries of the supercell (blue lines), and a dotted line drawn inbetween where the two dimer rows would appear on a reconstructed (001) silicon surface.}
	\label{fig:1-row}
\end{figure}

The \siesta~package was used to apply the Kohn-Sham self-consistent density-functional method in the generalized-gradient approximation~\cite{Kohn1965a,Ordejon1996a,Soler2002a}. This method was used to calculate the equilibrium electronic properties of the Si:P $\delta$-doped wire shown in Fig.~\ref{fig:1-row}{a}. This wire is made of a single row of phosphorus atoms that have been doped into one dimer row (1DR) of the reconstructed (001) silicon surface and will therefore be referred to as the \textit{single-row wire}. The single-row wire represents the 1D limit to scaling of the in-plane width of a phosphorus wire in silicon\footnote{At a 2D doping density of 0.25~ML.}. A supercell for the single-row wire is shown in Fig.~\ref{fig:1-row}{b} where the axis of the wire is oriented in the $\left[110\right]$ direction. This supercell has dimensions of $0.77~\mathrm{nm}\times5.40~\mathrm{nm}\times5.46~\mathrm{nm}$, which corresponds to at least 2.7~nm of bulk silicon ``cladding'' perpendicular to the wire axis\footnote{The supercell in Fig.~\ref{fig:1-row}{b} contains a total of 1120 atoms.}. To avoid surface effects in the calculations, periodic boundary conditions are applied to the supercells. In Fig.~\ref{fig:1-row}, the periodic boundaries are shown as blue lines. The length of the supercell in the $\left[1\bar{1}0\right]$ and $\left[001\right]$ directions is therefore determined by the amount of bulk silicon cladding that is needed to isolate the phosphorus wire from its periodic images.

\begin{figure}[t!]
    \centering
	\includegraphics[]{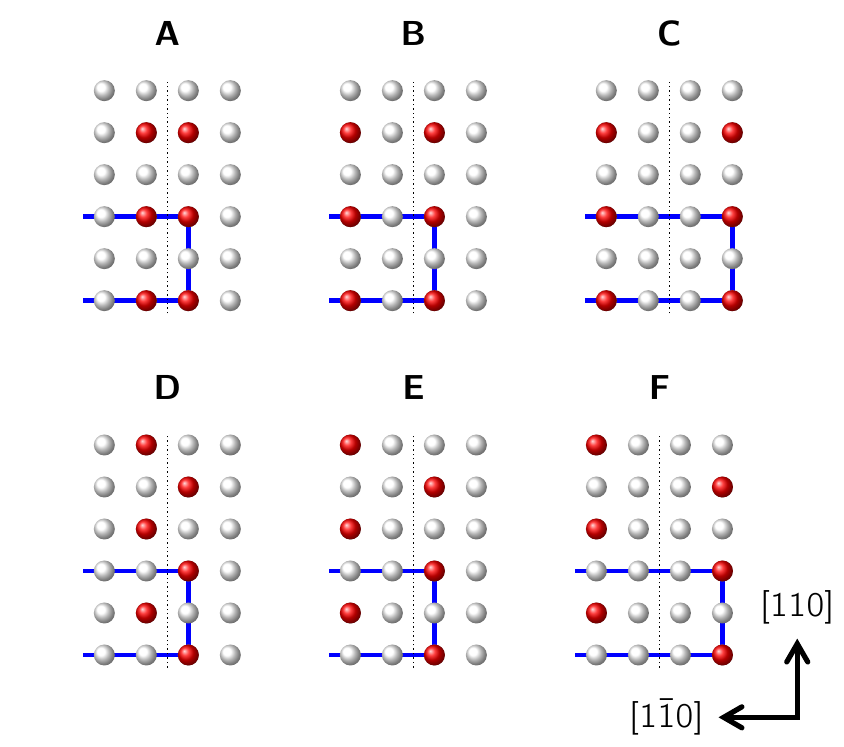}
	\caption{(color online) 2D schematics of double-row wires A-F (see labels) showing only part of the donor plane with silicon atoms (white spheres), phosphorus atoms (red spheres), the periodic boundaries of the supercell (blue lines), and a dotted line drawn inbetween where the two dimer rows would appear on a reconstructed (001) silicon surface. The periodic boundaries are drawn such that there is always a donor atom at the origin of the supercells.}
	\label{fig:2-row}
\end{figure}

We have also performed density-functional calculations on the Si:P $\delta$-doped wires shown in Fig.~\ref{fig:2-row}. These wires have been doped into two dimer rows (2DRs) of the reconstructed (001) silicon surface\cite{Note1} and will be referred to as \textit{double-row wires A-F} (see labels in Fig.~\ref{fig:2-row}). Double-row wires A-F represent all donor configurations which can result from adsorption of $\rm{PH}_{3}$ molecules onto a 2DR wide and 0.77~nm long region on the Si(001) surface, which is also periodic in the $\left[110\right]$ direction\cite{Note1}. For double-row wire B, we have previously found that by using 2.7~nm of bulk silicon cladding perpendicular to the wire axis, the minima of the occupied bands in the resulting band structure are converged to within 5~meV~\cite{Drumm2013a}.

To reduce the periodicity of the donor atoms in the $\left[110\right]$ direction we would need to at least double the length of the supercell in this direction\cite{Note1}. When the length of the supercell is doubled in one direction, the number of atoms in the supercell also doubles. This results in an impractical eightfold increase in the computation time of density-functional calculations. Therefore, the length of the supercells in the $\left[110\right]$ direction is limited to 0.77~nm for all wires considered herein. For double-row wires A-F, the lengths of the supercell in the $\left[110\right]$ and $\left[001\right]$ directions are the same as that for the supercell shown in Fig.~\ref{fig:1-row}{b}. However, the length of the supercell in the $\left[1\bar{1}0\right]$ direction is larger so that there is always at least 2.7~nm of bulk silicon cladding perpendicular to the wires.

\begin{figure}[b!]
    \centering
    \includegraphics[]{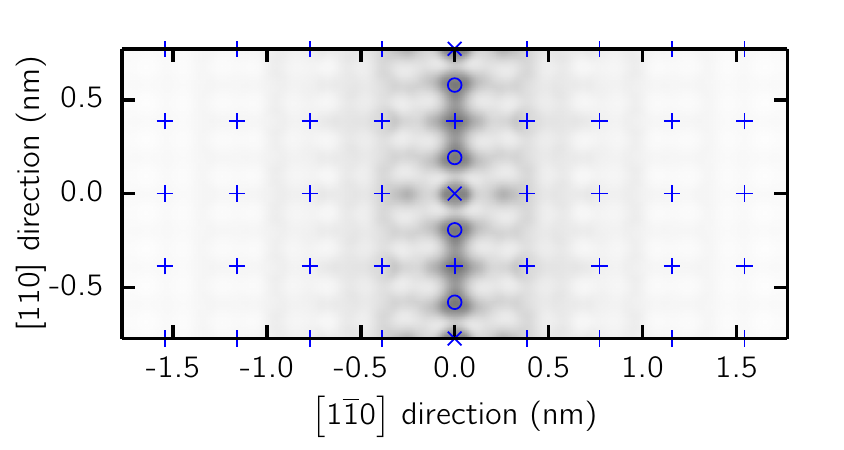}
    \caption{(color online) The local density of states for the single-row wire (integrated between $\Gamma_{1}$ and the equilibirum Fermi level and then line-averaged in the $\left[001\right]$ direction) showing the probability density of the donor electrons (dark) with marked positions for phosphorus atoms $\left(\times\right)$, out-of-plane silicon atoms along the wire axis $\left(\circ\right)$, and in-plane silicon atoms $\left(+\right)$.}
    \label{fig:1-row-ldos}
\end{figure}

An optimized double-$\zeta$ polarized basis set of localized atomic orbitals and norm-conserving Troullier-Martins pseudopotentials were used to variationally solve for the ground-state electron density of these $\delta$-doped wires~\cite{Budi2012a,Troullier1991a}. The lattice constant of bulk silicon was found to be 5.4575~\AA~using this basis set, which is in good agreement with the experimental value of 5.431~\AA~\cite{Becker1982a}. The exchange-correlation energy was calculated using the Perdew-Burke-Ernzerhof exchange-correlation functional~\cite{Perdew1996a}. Total energies were converged to within $10^{-4}~\mathrm{eV}$ using a Methfessel-Paxton occupation function of order 5 with an electronic smearing of 0.026~eV. The mesh energy grid cut-off used was 300~Ry. A $6\times1\times1$ Monkhorst-Pack $k$-point grid was used to sample the Brillouin zone (BZ), which has previously been shown to give good results for double-row wire B~\cite{Drumm2013a}. The atomic positions of the silicon atoms were not geometrically optimized after phosphorus substitution because their positions are not significantly affected by this substitution~\cite{Drumm2013b,Carter2009a}. Further details of our density-functional calculations are available in Refs.~\onlinecite{Budi2012a,Drumm2013a,Drumm2013b}.

\subsection{\label{sec:results-ldos}Probability density of donor electrons}

\begin{figure}[t!]
    \centering
    \includegraphics[]{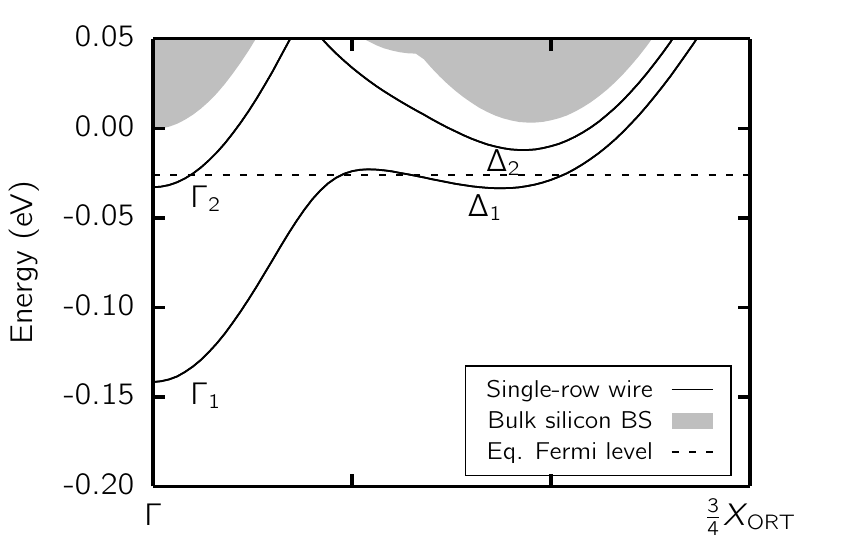}
    \caption{The band structure for the single-row wire between $\Gamma$ and $\frac{3}{4}X_{\mathrm{ORT}}$ (solid lines) with the equilibrium Fermi level (dashed line) and the band structure for bulk silicon (shaded region). The CB minimum of bulk silicon has been set to energy zero and the point $X_{\mathrm{ORT}}$ is at $\frac{1}{2\sqrt{2}}\frac{2\pi}{a}$ in the $\left[110\right]$ $k$-space direction. For means of comparison, this band structure has been calculated using a supercell with the same dimensions as the supercells of double-row wires A, B, D, and E.}
    \label{fig:1-row-bands}
\end{figure}

\begin{figure*}[t!]
    \centering
    \includegraphics[]{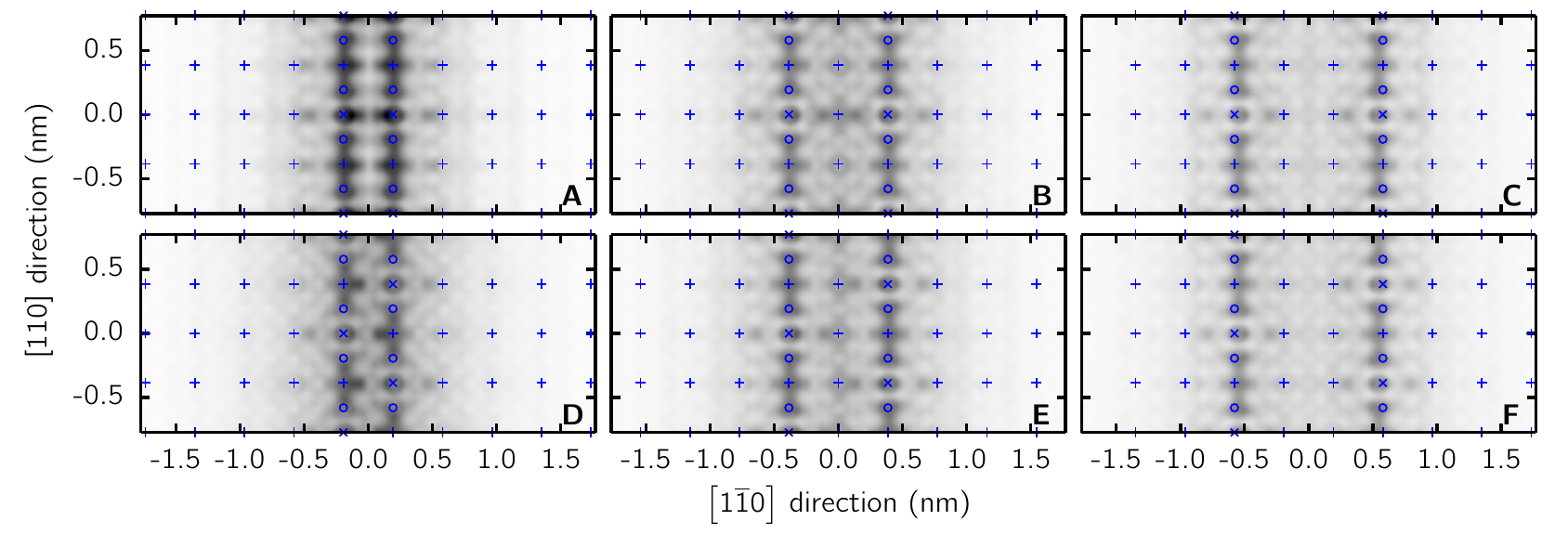}
    \caption{(color online) The local density of states for double-row wires A-F (integrated between the $\Gamma_{1}$ minimum and the equilibrium Fermi level and then line-averaged in the $\left[001\right]$ direction) showing the probability density of the donor electrons (dark) with marked positions for phosphorus atoms $\left(\times\right)$, out-of-plane silicon atoms along the wire axes $\left(\circ\right)$, and in-plane silicon atoms $\left(+\right)$.}
    \label{fig:2-row-ldos}
\end{figure*}

The probability density for the donor electrons in the single-row wire is shown in Fig.~\ref{fig:1-row-ldos}. This probability density has been calculated by integrating the local density of states (LDOS) between the conduction band (CB) edge and the equilibrium Fermi level of the single-row wire band structure shown in Fig.~\ref{fig:1-row-bands}. Therefore, this probability density is a mixture of the states with minima $\Gamma_{1}$, $\Gamma_{2}$, and $\Delta_{1}$ (which are labeled in Fig.~\ref{fig:1-row-bands}). We can assume the donor electrons occupy the bands in this energy range because the valence band of bulk silicon is fully occupied at equilibrium and so the only bands that are available for the donor electrons are those of the CB. Experimental measurements of these systems are performed at  $T=4.2~\mathrm{K}$ and so we can also ignore the effects of electronic smearing (due to thermal motion) in this calculation~\cite{Weber2012a,Weber2014a}. The LDOS is calculated for the full supercell on a three-dimensional (3D) Cartesian grid, which is then line-averaged along the [001] direction (perpendicular to the donor plane) to make the probability density shown in Fig.~\ref{fig:1-row-ldos}. The probability distribution is normalized such that the integral of this probability density over the supercell is equal to the number of donor electrons inside the supercell. In Fig.~\ref{fig:1-row-ldos}, a distance equivalent to two supercells is shown in the [110] direction.

The donor electrons are partially delocalized along the axis of the phosphorus wire in Fig.~\ref{fig:1-row-ldos}. There is significant probability density not only on the phosphorus atoms but also the intervening silicon atoms both in- and out-of-plane. The majority of the probability density is localized to the atomic sites along the axis of the wire. Fig.~\ref{fig:1-row-ldos} shows there is strong spatial confinement of the donor electrons perpendicular to the axis of the wire. The probability density decays sharply with distance from the wire axis in the $[1\bar{1}0]$ direction. The majority of the probability density can seen to be localized to $\pm0.5$~nm of the wire axis and, therefore, one may conclude that $\sim1$~nm is a good approximation for the electronic extent of the wire. However, as we will demonstrate, this is not a valid approximation. Indeed, we will show that even $\sim2$~nm is a poor approximation for the electronic extent of the single-row wire.

Fig.~\ref{fig:2-row-ldos} shows the probability densities for double-row wires A-F. These probability densities have been calculated in the same way as Fig.~\ref{fig:1-row-ldos}. In each subfigure, there is significant overlap between the wavefunctions of the two rows of phosphorus atoms. The presence of this overlap is independent of whether the rows are staggered relative to one another or aligned (compare the probability densities of double-row wires A-C with those of double-row wires D-F). If the overlap between the two rows of phosphorus atoms is large enough such that they behave as a single wire, then the electronic width of this wire is strongly dependent on the donor configuration. The electronic width varies by as much as $\sim1~\mathrm{nm}$ over double-row wires A-F and, therefore, we expect a similar variation in the width of a realistic wire.

Overall, it is difficult to determine the electronic extent of a $\delta$-doped wire from these probability densities. Instead, we use the band structure of the wires and, in particular, energy splittings in the band minima to determine the electronic extent of a $\delta$-doped wire.

\subsection{\label{sec:results-bands}Band structure of $\delta$-doped wires}

The band structure of bulk silicon calculated using a 1280-atom orthorhombic (ORT) supercell is shown as the gray shaded region in Fig.~\ref{fig:1-row-bands}. There are two CB edges shown for bulk silicon, which are each doubly degenerate; one at $\Gamma$ and the other at $0.46X_{\mathrm{ORT}}$, where $X_{\mathrm{ORT}}=\frac{1}{2\sqrt{2}}\frac{2\pi}{a}$ in the $\left[110\right]$ $k$-space direction. An explanation of the location of the bulk CB edges in this band structure is given in Appendix~\ref{sec:appendix}. The band structure of the single-row wire is shown as solid lines in Fig.~\ref{fig:1-row-bands}. The nuclear potential of the phosphorus atoms has moved the CB edges into the bulk bandgap region. The degeneracy of these CB edges or valleys has been broken, resulting in the appearance of four separate valleys. The minima of these four valleys are labeled as $\Gamma_{1}$, $\Gamma_{2}$, $\Delta_{1}$, and $\Delta_{2}$ in Fig.~\ref{fig:1-row-bands}.

\begin{figure*}[t!]
    \centering
    \includegraphics[]{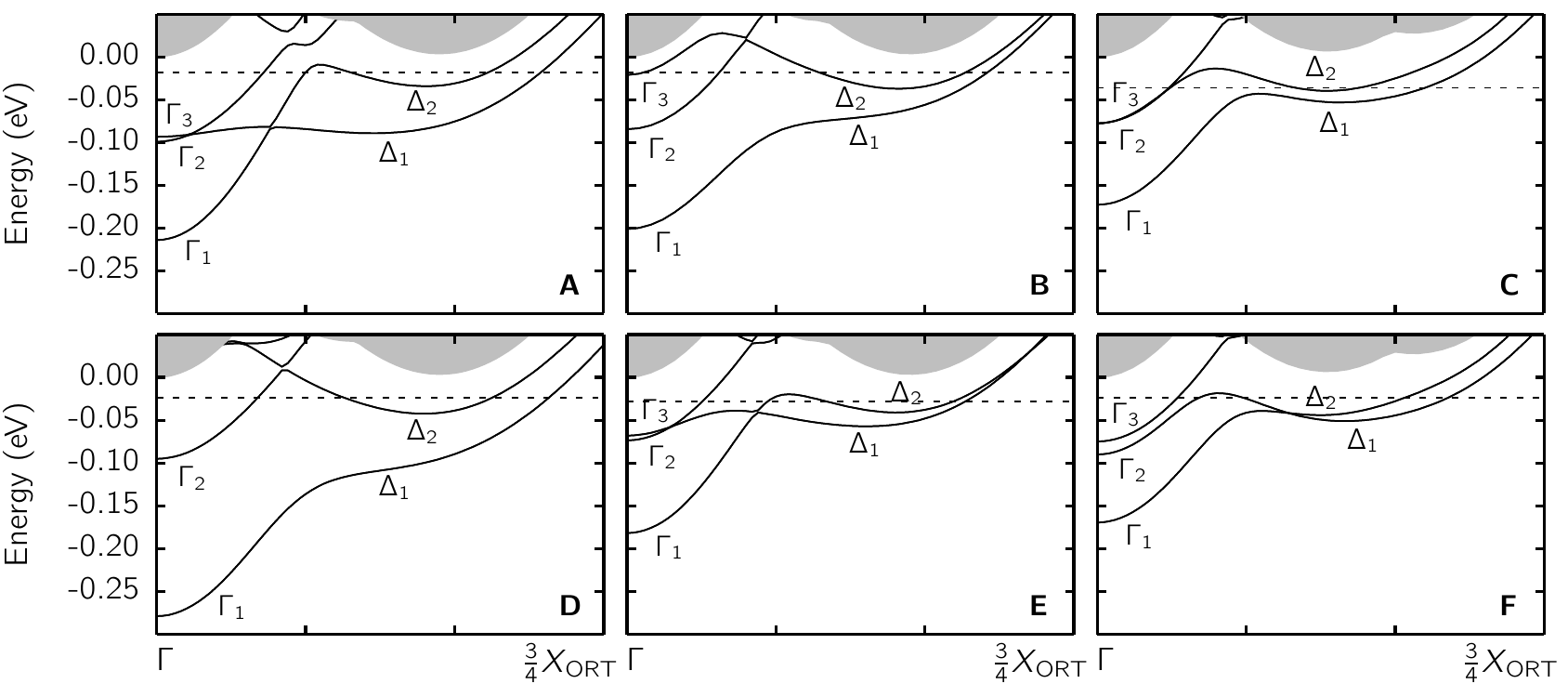}
    \caption{The band structure for double-row wires A-F between $\Gamma$ and $\frac{3}{4}X_{\mathrm{ORT}}$ (solid lines) with the equilibrium Fermi level (dashed line) and the band structure for bulk silicon (shaded region). The CB minimum of bulk silicon has been set to energy zero and the point $X_{\mathrm{ORT}}$ is at $\frac{1}{2\sqrt{2}}\frac{2\pi}{a}$ in the $\left[110\right]$ $k$-space direction. For double-row wires C and F, the location of the bulk silicon CB minimum along the path $\Gamma\to X_{\mathrm{ORT}}$ has changed because the length of the supercell in the $\left[1\bar{1}0\right]$ direction is larger for these wires (see Appendix~\ref{sec:appendix}).}
    \label{fig:2-row-bands}
\end{figure*}

It is well-known for $\delta$-doped systems that an enhancement of the valley-orbit interaction due to spatial confinement will break the six-fold degeneracy of the bulk silicon CB edge, resulting in a valley splitting (VS)~\cite{Carter2009a,Carter2011a,Drumm2012a,Drumm2013b}. In Fig.~\ref{fig:1-row-bands}, the 1D confinement caused by the donor potential has moved the CB edges into the bulk bandgap region and lifted their degeneracy by a VS. We report two different VSs for the single-row wire; one at $\Gamma$ labeled $\Gamma_{1}-\Gamma_{2}$ and another at $k=0.46X_{\mathrm{ORT}}$ labeled $\Delta_{1}-\Delta_{2}$. The magnitude of the $\Gamma_{1}-\Gamma_{2}$ splitting is found to be 109~meV, which is within 10~meV of the same VS reported for $\delta$-doped layers~\cite{Drumm2013b}. It has also been reported for $\delta$-doped layers that CB valleys with higher curvature are moved further into the bulk bandgap region~\cite{Drumm2012a,Smith2014a} and this is shown for the single-row wire in Fig.~\ref{fig:1-row-bands}. In addition, we find the VS is larger for CB valleys with higher curvature. The magnitude of the $\Delta_{1}-\Delta_{2}$ splitting is found to be 21~meV, which is 19\% of the $\Gamma_{1}-\Gamma_{2}$ splitting.

The position of the Fermi level shows three of the four CB valleys to be occupied at equilibrium; the $\Gamma_{1}$, $\Gamma_{2}$, and $\Delta_{1}$ bands. The $\Delta_{1}$ and $\Delta_{2}$ minima at $k=0.46X_{\mathrm{ORT}}$ in the $\left[110\right]$ $k$-space direction are symmetrically equivalent to two CB edges at $k=-0.46X_{\mathrm{ORT}}$. This is due to the symmetry of the path $\Gamma\to X_{\mathrm{ORT}}$ which is discussed in Appendix~\ref{sec:appendix}. Therefore, for the single-row wire, we predict four conducting modes to be available for electron transport at low voltage biases.

The band structures for double-row wires A-F are shown in Fig.~\ref{fig:2-row-bands}. The location of the bulk silicon CB edge for double-row wires C and F is at $k\approx0.38X_{\mathrm{ORT}}$ not $k\approx0.46X_{\mathrm{ORT}}$, as discussed in Appendix~\ref{sec:appendix}. For the double-row wires, the larger nuclear potential of more phosphorus atoms moves the CB valleys further into the bulk bandgap region compared to the single-row wire. The minima of these valleys are again labeled as $\Gamma_{1}$, $\Gamma_{2}$, $\Delta_{1}$, and $\Delta_{2}$. For all double-row wires (except double-row wire D) there is a third CB valley minimum at the $\Gamma$ point, labeled as $\Gamma_{3}$ in Fig.~\ref{fig:2-row-bands}. The position of the Fermi level shows five CB valleys (including $\Gamma_{3}$) to be occupied at equilibrium for double-row wires A, C, E, and F. We expect there to be eight available conducting modes for the double-row wires, which is twice the number of available conducting modes as the single-row wire. However, there are only seven available conducting modes for double-row wires A, C, E, and F and six for double-row wires B and D when the symmetry of the path $\Gamma\to X_{\mathrm{ORT}}$ is taken into account.

The magnitudes of the $\Gamma_{1}-\Gamma_{2}$ and $\Delta_{1}-\Delta_{2}$ splittings are different for each of the double-row wires and, therefore, these VSs must be dependent on donor configuration. It has previously been reported for $\delta$-doped layers that the magnitude of the $\Gamma_{1}-\Gamma_{2}$ splitting is dependent on the in-plane configuration of the donor atoms~\cite{Carter2011a}. Fig.~\ref{fig:2-row-bands} shows the magnitude of the VS is affected by whether the two rows of phosphorus atoms are aligned (A, B, C) or staggered (D, E, F) relative to one another. The $\Gamma_{1}-\Gamma_{2}$ and $\Delta_{1}-\Delta_{2}$ splittings are largest for double-row wires A and D. These wires represent the donor configurations for which the two rows of phosphorus atoms are laterally separated by the smallest distance. The $\Gamma_{1}-\Gamma_{2}$ and $\Delta_{1}-\Delta_{2}$ splittings decrease as the distance of lateral separation increases in Fig.~\ref{fig:2-row-bands}.

\subsection{\label{sec:results-overlap}Electronic extent of a $\delta$-doped wire in the one-dimensional limit}

We expect there to be eight available conducting modes for double-row wires A-F. However, in the previous subsection, we report seven (and six) available conducting modes for double-row wires A, C, E, and F (and double-row wires B and D). Therefore, the number of available conducting modes in these wires is reduced by something so far unaccounted for. If there were zero overlap between the wavefunctions of each row of phosphorus atoms, the band structure for the double-row wires would be identical to the band structure of the single-row wire (except with each band being doubly degenerate). Therefore, we suggest $\Gamma_{3}$ is a degenerate pair of $\Gamma_{1}$ and that this degeneracy has been lifted by an energy splitting which is proportional to the wavefunction overlap between the two rows of phosphorus atoms\footnote{Given this hypothesis, we suggest the $\Gamma_{1}$-$\Gamma_{3}$ splitting is not observed for double-row D because the splitting is so large that the $\Gamma_{3}$ minima cannot be distinguished from the higher energy modes of the silicon band structure.}. We label this energy splitting as $\Gamma_{1}$-$\Gamma_{3}$. In Fig.~\ref{fig:2-row-bands}, excluding double-row wires A and D, the $\Gamma_{1}$-$\Gamma_{3}$ splitting decreases as the lateral separation of the two rows increases, \textit{i.e.} as the wavefunction overlap between the two rows decreases so too does the $\Gamma_{1}$-$\Gamma_{3}$ splitting. A similar energy splitting has previously been reported for two adjacent $\delta$-doped layers~\cite{Carter2013a,Drumm2014a}.

\begin{figure}[t!]
    \centering
    \includegraphics[]{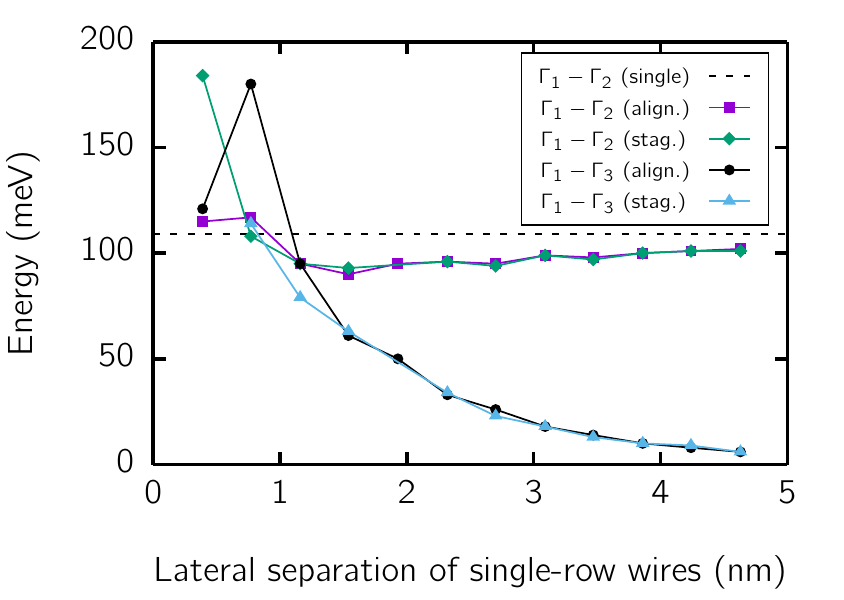}
    \caption{(color online) The $\Gamma_{1}-\Gamma_{2}$ and $\Gamma_{1}-\Gamma_{3}$ splittings versus the lateral separation of two single-row wires when they are aligned and staggered with respect to each other. The $\Gamma_{1}-\Gamma_{2}$ splitting for an isolated single-row wire is also shown (dashed line).}
    \label{fig:splittings}
\end{figure}

In Fig.~\ref{fig:splittings}, the $\Gamma_{1}$-$\Gamma_{3}$ splitting is plotted versus the lateral separation of the two rows of phosphorus atoms (between 0.8~nm and 4.6~nm). We also plot the $\Gamma_{1}$-$\Gamma_{2}$ splitting versus the lateral separation of the two rows of phosphorus atoms in this figure. These energy splittings have been calculated by using supercells that are larger than those used for double-row wires A-F\footnote{The lengths of these supercells in the $\left[110\right]$ and $\left[001\right]$ directions are the same as that for double-row wires A-F. However, the length of the supercells in the $\left[1\bar{1}0\right]$ direction is larger. In this direction, the length of the supercells is chosen such that there is always at least 2.7~nm of bulk silicon cladding surrounding the two rows of phosphorus atoms. Therefore, the maximum distance of lateral separation that can be investigated has to be less than 5.4~nm.}. In Fig.~\ref{fig:splittings}, the $\Gamma_{1}$-$\Gamma_{2}$ splitting tends towards the value of the VS calculated for the single-row wire as the lateral separation of the two rows is increased. By contrast, the $\Gamma_{1}$-$\Gamma_{3}$ splitting tends towards zero as the lateral separation is increased. This suggests $\Gamma_{3}$ is a degenerate pair of $\Gamma_{1}$ and that the wavefunction overlap between the two rows of phosphorus atoms is what breaks this degeneracy.

At a lateral separation of 1.9~nm, two additional CB valley minima appear at $k\approx0.46X_{\mathrm{ORT}}$ (when the two rows of phosphorus atoms are aligned), which we interpret as degenerate pairs of $\Delta_{1}$ and $\Delta_{2}$ that have also been lifted by an energy splitting. These splittings tend to zero as the lateral separation of the two rows is further increased. When the lateral separation is equal to 3.5~nm, a fourth CB valley minimum appears at $\Gamma$ which we suggest is the degenerate pair of $\Gamma_{2}$. The energy splitting of the $\Gamma_{2}$ degeneracy is equal to 60~meV at a lateral separation of 3.5~nm and decreases to 37~meV by 4.6~nm, \textit{i.e.} it has not converged to zero at a lateral separation of 4.6~nm.

We can now use this analysis of the $\Gamma_{1}-\Gamma_{3}$ splitting (and other energy splittings in $\Gamma_{2}$, $\Delta_{1}$, and $\Delta_{2}$) to approximate the electronic extent of the single-row wire. When the lateral separation of the two rows of phosphorus atoms is large, the two rows will each behave as single-row wires. The electronic extent of the single-row wire is then the lateral separation at which the energy splittings in $\Gamma_{1}$, $\Gamma_{2}$, $\Delta_{1}$, and $\Delta_{2}$ are equal to zero. The $\Gamma_{1}$-$\Gamma_{3}$ splitting and the energy splittings of the $\Delta_{1}$ and $\Delta_{2}$ degeneracies are approximately equal to 6~meV at a lateral separation of 4.6~nm. This is equal to the uncertainty in $\Gamma_{1}$, $\Gamma_{2}$, $\Delta_{1}$ and $\Delta_{2}$. Therefore, to within the uncertainty of our density-functional method, the energy splittings in $\Gamma_{1}$, $\Delta_{1}$ and $\Delta_{2}$ are indistinguishable from zero and so the electronic extent of the single-row wire is approximately equal to 4.6~nm. However, because the energy splitting in the $\Gamma_{2}$ degeneracy has not converged to zero by 4.6~nm, this is but a lower bound on the electronic extent of the single-row wire. Nonetheless, we expect this to be a good approximation because the occupancy of the $\Gamma_{1}$ state is much greater than that of the $\Gamma_{2}$ state, as shown in Fig.~\ref{fig:1-row-bands}. For $\delta$-doped wires that are separated by in-plane distances less than 4.6~nm, we predict a decrease in the number of available conducting modes at low voltage biases.

\section{\label{sec:negf}Electronic transport properties of $\delta$-doped wires}

In this section we use the results of our density-functional calculations to construct a computational model of electron transport in a Si:P $\delta$-doped wire. We solve for the electronic transport properties of a $\delta$-doped wire using the general NEGF approach of Datta and others~\cite{Datta2006a,Golizadeh-Mojarad2007a,Svizhenko2007a,Ryndyk2009a}.

\subsection{\label{sec:methods-negf}The nonequilibrium Green's function formalism}

The difficulty of solving for the eigenstates of a many-body system in the Schr\"odinger picture is avoided in the NEGF formalism by replacing the Hamiltonian operator by a Green's function matrix. The transport properties of the system are then calculated from this Green's function matrix. Our NEGF model describes a Si:P $\delta$-doped wire using a TB Hamiltonian matrix, within a single-band effective-mass approximation, that is defined as
\begin{equation}
    \mathbf{H}=\sum_{i}^{N}\ \varepsilon\ \ket{i}\bra{i}-\sum_{i,j,i\neq j}^{N}\ t\ \ket{i}\bra{j}
    \label{eq:hamiltonian}
\end{equation}
where $\varepsilon=-2Dt+U$ is the on-site energy, $t=\hbar^2/2\bar{m}\alpha^2$ is the tunneling parameter, $i$ and $j$ are first-nearest-neighbor donor atoms, and $N$ is the total number of donor atoms. $U$ is an offset to the on-site energy due to a gate voltage applied to the wire (with $U=0$~eV for a gate voltage equal to zero), $D$ is the dimension of the device, $\bar{m}$ is the effective mass of the donor electrons (discussed below), and $\alpha$ is the distance between two nearest-neighbor donor atoms.

A Si:P $\delta$-doped wire is divided into three parts: a source, a drain, and a channel region that separates the two. In general, Eq.~\ref{eq:hamiltonian} describes the channel region only. However, in our NEGF model the source and drain contacts are described as semi-infinite extensions of the channel and, therefore, Eq.~\ref{eq:hamiltonian} is also used to describe the contacts. The retarded Green's function matrix is then defined as
\begin{equation}
    \mathbf{G}\left(E\right)=\left[\left(E+i\eta\right)\mathbf{I}-\mathbf{H}-\mathbf{\Sigma}_{\mathrm{S}}-\mathbf{\Sigma}_{\mathrm{D}}\right]^{-1}
    \label{eq:greensfunction}
\end{equation}
where $E$ is energy, $\eta$ is a positive infinitesimal real number, and $\mathbf{\Sigma}_{\mathrm{S}}$ and $\mathbf{\Sigma}_{\mathrm{D}}$ are the self-energy matrices for the source and drain respectively. These are given by
\begin{equation}
    \mathbf{\Sigma}_{\mathrm{S,D}}={\boldsymbol\tau}^{\phantom{\dagger}}_{\mathrm{S,D}}\mathbf{g}^{\phantom{\dagger}}_{\mathrm{S,D}}{\boldsymbol\tau}^{\dagger}_{\mathrm{S,D}}
\end{equation}
where $\boldsymbol\tau$ is the coupling matrix between the contact and the channel, and $\mathbf{g}$ is the surface Green's function for the contact. We calculate the surface Green's functions using the iterative scheme of \citeauthor{Sancho1985a}, which solves the accompanying Dyson equation to arbitrary precision~\cite{Sancho1985a}.

The transmission function for the device is written as
\begin{equation}
    T\left(E\right)=d\times\mathrm{tr}(\mathbf{\Gamma}_{\mathrm{S}}\mathbf{G}\mathbf{\Gamma}_{\mathrm{D}}\mathbf{G}^{\dagger})
    \label{eq:transmission}
\end{equation}
where $d$ is the degeneracy of the single band (discussed below), and $\mathbf{\Gamma}_{\mathrm{S}}$ and $\mathbf{\Gamma}_{\mathrm{D}}$ are the broadening matrices for the source and the drain respectively. These are given by
\begin{equation}
    \mathbf{\Gamma}_{\mathrm{S,D}}=i\left[\mathbf{\Sigma}_{\mathrm{S,D}}-\left(\mathbf{\Sigma}_{\mathrm{S,D}}\right)^{\dagger}\right]
\end{equation}
In the Landauer-B\"uttiker formalism, the current can then be calculated using the equation:
\begin{equation}
    I=\frac{q}{h}\int_{-\infty}^{+\infty}T\left(E\right)\left[f_{\mathrm{S}}\left(E\right)-f_{\mathrm{D}}\left(E\right)\right]dE
\end{equation}
where $q$ is the elementary charge of an electron, $h$ is Planck's constant and $f\left(E\right)$ is the Fermi-Dirac distribution function for the contacts, defined as
\begin{equation}
    f_{\mathrm{S,D}}\left(E\right)=\frac{1}{1+e^{\left(E-\mu_{\mathrm{S,D}}\right)/k_{\mathrm{B}}T}}
    \label{eq:fermi}
\end{equation}
In Eq.~\ref{eq:fermi}, $k_{\mathrm{B}}$ is Boltzmann's constant, $T$ is temperature (in Kelvin), and $\mu_{\mathrm{S,D}}$ is the chemical potential of the source or drain which are given by
\begin{equation}
    \mu_{\mathrm{S}} = \mu+\frac{V_{\mathrm{SD}}}{2}
\end{equation}
and
\begin{equation}
    \mu_{\mathrm{D}} = \mu-\frac{V_{\mathrm{SD}}}{2}
\end{equation}
with $\mu$ the equilibrium chemical potential of the wire\footnote{The equilibrium chemical potential~$\mu$ is measured relative to the energy of the CB edge when $U=0$~eV, which is not necessarily energy zero.} and $V_{\mathrm{S,D}}$ the source-drain bias voltage. For our calculations, the source-drain bias voltage decays linearly across the channel region.

The effective mass of the donor electrons is needed to fully specify the TB Hamiltonian in Eq.~\ref{eq:greensfunction}. This effective mass can be calculated for the single-row wire from the occupied CB valleys in the band struture shown in Fig.~\ref{fig:1-row-bands}. The band dispersion in the neighborhood of the CB valley minima is approximately parabolic~\cite{Drumm2012a,Drumm2013a}. The curvature $\beta$ of this parabola is related to the effective mass $\bar{m}$ of the donor electrons through the equation~\cite{Drumm2013a}:
\begin{equation}
    \frac{\hbar^{2}k^{2}}{2\bar{m}}=\beta k^{2}
\end{equation}
Therefore, the effective mass of the donor electrons can be calculated by fitting parabola to the CB valleys in Fig.~\ref{fig:1-row-bands}. However, it has previously been shown for double-row wire B that the curvature of these bands do not change significantly from their bulk values and so we may use the transverse and longitudinal effective masses of bulk silicon without modification~\cite{Drumm2013a}. In Fig.~\ref{fig:1-row-bands}, the CB valleys at $\Gamma$ have high curvature; they are described by the transverse effective mass $m_{t}$ of bulk silicon~\cite{Drumm2013a}. The CB valleys at $|k|\approx0.46X_{\mathrm{ORT}}$ have low curvature; they are described by the longitudinal effective mass $m_{l}$ of bulk silicon~\cite{Drumm2013a}.

\begingroup
\begin{table}[t!]
\begin{ruledtabular}
     \caption{The values used for the free parameters in our NEGF model. $m_{0}$ is the free electron mass.}
     \begin{tabular*}{\columnwidth}{@{\extracolsep{\fill} } c c c c c c }
          $m_{l}/m_{0}$ & $m_{t}/m_{0}$ & $\alpha$~(\AA) & $T$ (K) & $\mu$ (eV) & $d$ \\
          \hline
          0.9163~\footnote{\label{Hensel1965a}Ref.~\onlinecite{Hensel1965a}} & $0.1905^{~\mathrm{a}}$ & 7.718 & 4.2~\footnote{Ref.~\onlinecite{Weber2014a}} & 0.12 & 4
     \label{tab:parameters}
     \end{tabular*}
\end{ruledtabular}
\end{table}
\endgroup

In Section~\ref{sec:dft}, it was shown for the single-row wire that there were four conducting modes available for electronic transport at low voltage biases. Therefore, we describe the single-row wire by one conducting mode that is four-fold degenerate within a single-band effective-mass approximation by setting $d=4$ in Eq.~\ref{eq:transmission}. We set the effective mass of this conducting mode equal to the average of the effective masses for the four occupied conducting modes. There are two occupied CB edges at $\Gamma$ and two occupied CB edges at $|k|\approx0.46X_{\mathrm{ORT}}$. The average of the effective masses is given by
\begin{equation}
    \bar{m}=\left(2m_{t}+2\frac{m_{l}}{2}\right)/4
\end{equation}
where there is a reduction of $\frac{1}{2}$ in the longitudinal effective mass (as discussed in Ref.~\onlinecite{Drumm2013a} and Appendix~\ref{sec:appendix}). The bulk silicon effective masses $m_{l}$ and $m_{t}$, and the other parameters used in our NEGF model, are listed in Table~\ref{tab:parameters}. The equilibrium chemical potential $\mu$ is equal to the difference between $\Gamma_{1}$ and the equilibrium Fermi level for the single-row wire in Fig.~\ref{fig:1-row-bands}. The atomic spacing $\alpha$ is equal to the distance between two nearest-neighbor donor atoms in Fig.~\ref{fig:1-row}{a}. 2D wires with widths greater than that of the single-row wire are modeled as many adjacent single-row wires where the lateral separation of the donor atoms is equal to $\alpha$ and the tunneling parameter for nearest-neighbor atoms in adjacent wires is equal to $t$. Hardwall boundary conditions are applied in the dimension perpendicular to the axes of the wires.

\subsection{\label{sec:results-negf}I-V characteristics of $\delta$-doped wires}

In this section we present I-V characteristics for two $\delta$-doped wires that have recently been measured in experiment~\cite{Weber2014a}. One of the wires has a width of 4.6~nm, which is equivalent to 6 dimer rows (6DRs) on the reconstructed (001) silicon surface. The other wire has a width of 1.5~nm, which is equivalent to 2DRs on the same surface. In addition, we present I-V characteristics for wires that have widths larger than 4.6~nm. For means of comparison, all the wires have a channel region with a length of 47~nm (which is equal to the length spanned by the 6DR wire in experiment~\cite{Weber2014a}).

\begin{figure}[t!]
    \centering
    \includegraphics[]{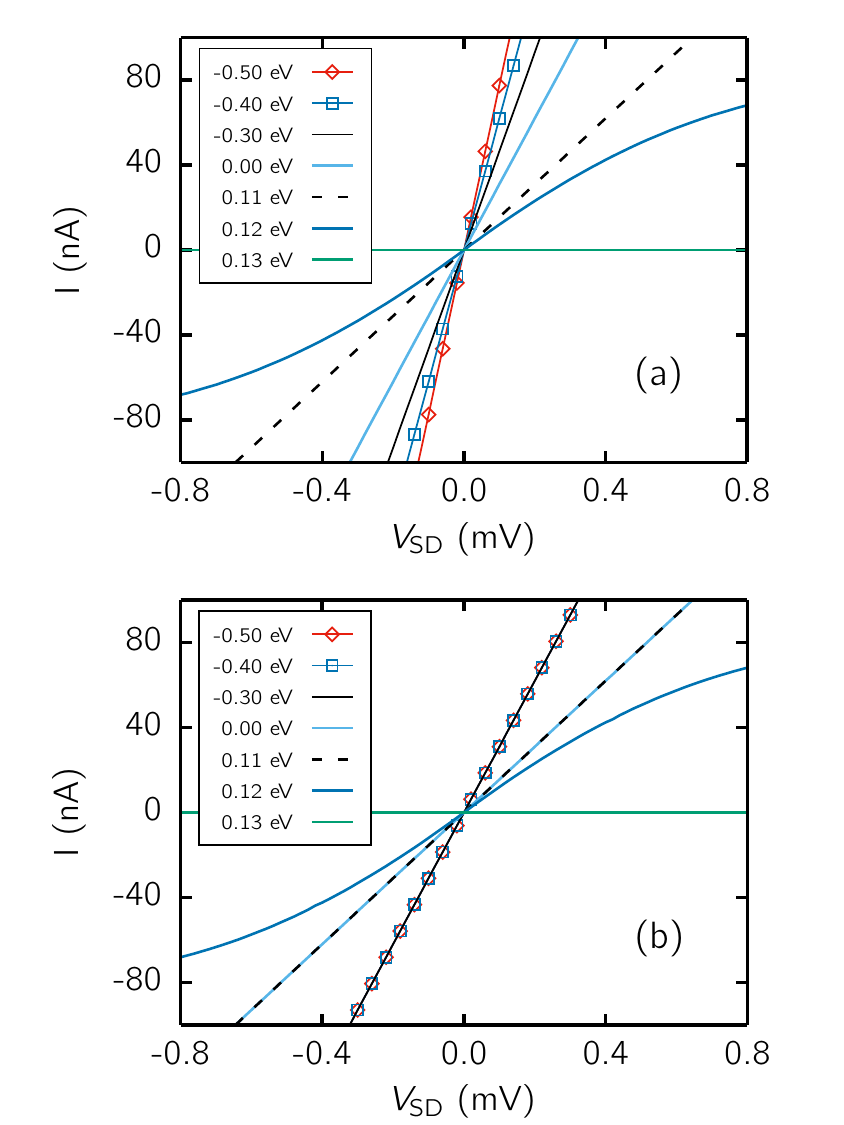}
    \caption{(color online) Current versus source-drain bias voltage for the (a) 6DR and (b) 2DR wire over a range of offset energies $U$. A negative offset energy is equivalent to a positive gate voltage in experiment.}
    \label{fig:IVchannelbias}
\end{figure}

The I-V characteristics of the 6DR wire are shown in Fig.~\ref{fig:IVchannelbias}{a}. The I-V curve for $U=0$~eV shows a linear response when a non-zero source-drain bias voltage is applied to the 6DR wire. This linear response is characteristic of metallic conduction and is in good agreement with experiment\cite{Weber2014a}. The size of the current is approximately 2 times greater than in experiment when $U=0$~eV (see Fig.~1e of Ref.~\onlinecite{Weber2014a}).

\begin{figure}[t!]
    \includegraphics[]{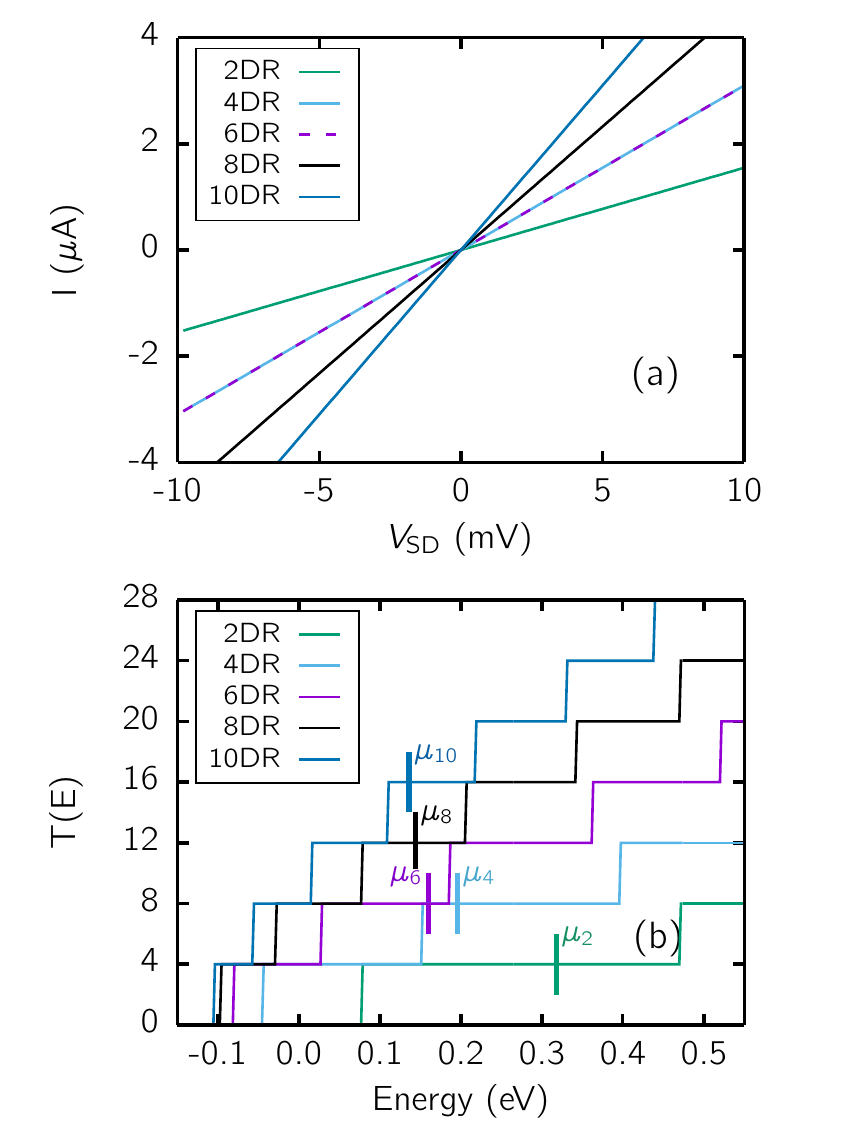}
    \caption{(color online) (a) Current versus source-drain voltage bias for wires with a variety of in-plane widths when $U=-0.12$~eV. (b) The transmission function $T(E)$ versus energy $E$ for each of these wires showing the position of the equilibrium chemical potential $\mu$ when $U=-0.12$~eV.} 
    \label{fig:IVwidth}
\end{figure}

In Fig.~\ref{fig:IVchannelbias}{b}, the I-V curve for $U=0$~eV shows a linear response when a bias voltage is applied to the 2DR wire. The size of the current is approximately 6 times greater than in experiment when $U=0$~eV (see Fig.~1f of Ref.~\onlinecite{Weber2014a}) and is exactly half that of the 6DR wire. Therefore, when $U=0$~eV there are double the number of available conducting modes in the 6DR wire than in the 2DR wire. The ratio of the two currents is also equal to two when $U=-0.12$~eV as shown in Fig.~\ref{fig:IVwidth}{a}. The doubling in the number of available conducting modes is confirmed by the transmission functions $T(E)$ for the two wires in Fig.~\ref{fig:IVwidth}{b} (compare the transmission function of the 6DR wire with that of the 2DR wire at $E=\mu$ in Fig.~\ref{fig:IVwidth}{b}).

The gradient of these I-V curves is equal to the differential conductance $G=dI/dV_{\mathrm{SD}}$ of the wires, which when divided by $e^2/h$ is equal to the number of conducting modes that are available for electron transport. To model the application of a gate voltage to the wires, we add a non-zero offset energy $U$ to the diagonal terms of the Hamiltonian matrix~\cite{Graf1995a} describing the channel region (see Eq.~\ref{eq:hamiltonian}). An applied gate voltage will change the occupancy of the conducting modes in the wire and thereby change the conductance of the device.

Beginning with the I-V curve for $U=0.13$~eV in Fig.~\ref{fig:IVchannelbias}{a}, the differential conductance of the 6DR wire increases as $U$ is decreased to zero and then becomes increasingly negative. A negative offset energy in our NEGF model is equivalent to a positive gate voltage in experiment and, therefore, this trend is in good agreement with experiment~\cite{Weber2014a} (where the differential conductance increases as increasingly positive gate voltages are applied to the 6DR wire [see Fig.~1e of Ref.~\onlinecite{Weber2014a}]). A negative offset energy moves the unoccupied conducting modes to lower energies, making them available to conduction electrons and thereby increasing the conductance of the device. Alternatively, a positive offset energy moves the occupied conducting modes to higher energies, emptying these states of electrons and resulting in zero current because there are no longer any states available to conduction electrons (as shown in Fig.~\ref{fig:IVchannelbias}{a} for $U=0.13$~eV).

I-V curves for the 2DR wire are shown in Fig.~\ref{fig:IVchannelbias}{b} for the same offset energies that are applied to the 6DR wire in Fig.~\ref{fig:IVchannelbias}{a}. The relationship between these offset energies and the differential conductance of the 2DR wire is the same as that of the 6DR wire up to $U=0.11$~eV. Beyond this point, the differential conductance for the 2DR wire increases when $U=-0.30$~eV and then remains constant as $U$ is made more negative (as shown for $U=-0.40$~eV and $U=-0.50$~eV). Therefore, when $U=-0.30$~eV, the current in the 2DR wire has reached saturation because all of the conducting modes are occupied. By contrast, in the wider 6DR wire there remain conducting modes at higher energies that can be made available to conduction electrons by applying larger gate voltages.

Finally, in both Figs.~\ref{fig:IVchannelbias}{a} and \ref{fig:IVchannelbias}{b}, there is a non-linear I-V response for $U=0.12$~eV (\textit{i.e.} for $U=\mu$). When the offset energy is equal to the equilibrium chemical potential, the mean energy of the conduction electrons is equal to the energy of the CB edge. The application of a source-drain voltage bias broadens the energy of the conduction electrons but it also increases the energy of the CB edge. And, therefore, it is possible for the rate of change in the energy of the CB edge to be greater than the rate of change in the energy broadening of the conduction electrons. If this is the case then although the current will increase as the source-drain bias voltage is increased, the change in the current will decrease (as shown in Figs.~\ref{fig:IVchannelbias}{a} and \ref{fig:IVchannelbias}{b} for $U=0.12$~eV). This non-linearity is not the same as the non-linear response reported experimentally for the 2DR wire (see Fig.~1f of Ref.~\onlinecite{Weber2014a}).

In general, there is good agreement between experiment and our results for the I-V characteristics of the 6DR wire. It is obvious that the results for the 6DR wire are in better agreement with experiment than the 2DR wire. This is the case for both the size of the current and the change in the current versus source-drain bias voltage. Therefore, we conclude that the 6DR wire is well-described by a ballistic model of electron transport, whereas the narrower 2DR wire is not. It is likely that a single-band effective-mass approximation is able to reproduce the low-temperature transport properties of the 6DR wire because the electron transport occurs at low voltage biases and, therefore, the higher energy modes of the silicon band structure are insignificant. This is promising for device simulation because our NEGF model scales easily to large system sizes.

There are a number of approximations in our NEGF model that could explain the discrepancies between theory and experiment. For the 6DR wire these include the value of the free parameters in Table~\ref{tab:parameters}, the boundary conditions, the average of the effective masses and the degeneracy of the single band in our effective-mass approximation. However, changing these properties will never reproduce the non-linear response reported experimentally for the 2DR wire~\cite{Weber2014a}. To reproduce this non-linearity, we suggest it is necessary to extend our NEGF model to include donor disorder and non-ballistic transport. In this paper we have chosen to present the simplest model of electron transport in a Si:P $\delta$-doped wire and, therefore, leave the investigation of these approximations to the subject of future work.

The transmission functions and equilibrium chemical potentials for wires with widths of 2DR, 4DR, 6DR, 8DR, and 10DR are shown in Fig.~\ref{fig:IVwidth}{b}. In this figure, $U$ is constant across all wires and is equal to -0.12~eV. The transmission functions and values of the equilibrium chemical potentials for each of these wires show that the number of available conducting modes increase as the width of the wires increase. Therefore, the I-V characteristics for the wider wires in Fig.~\ref{fig:IVwidth}{a} have larger differential conductances than the narrower wires. The energy of the lowest conducting modes decrease as the width of the wires increase, as shown in Fig.~\ref{fig:IVwidth}{b}. The spatial confinement of the donor electrons, perpendicular to the wire axis, decreases as the width of the $\delta$-doped wire increases; thereby moving the conducting modes to lower energies in a similar fashion to the eigenenergies of an infinite potential well. This decrease in spatial confinement also decreases the energy splittings between the conducting modes, which is shown in Fig.~\ref{fig:IVwidth}{b} as a narrowing of the steps in the transmission functions. Finally, it should be noted that the width of the steps in the transmission functions are related to the $\Gamma_{1}-\Gamma_{3}$ splitting\footnote{As well as the other energy splittings in the $\Gamma_{2}$, $\Delta_{1}$, and $\Delta_{2}$ minima that were discussed in Section~\ref{sec:dft}, excluding the VSs.}. Otherwise the transmission functions for the 2DR, 4DR, 6DR, 8DR, and 10DR wires would vary only by a multiplicative constant.

\section{\label{sec:conclusions}Conclusions}

The band structure of a $\delta$-doped wire has been calculated for a variety of donor configurations. We find a valley splitting at $\Gamma$, in agreement with previous density-functional calculations of $\delta$-doped layers. In addition, for $\delta$-doped wires comprised of more than a single row of phosphorus atoms, we find another energy splitting at $\Gamma$. This energy splitting ($\Gamma_{1}-\Gamma_{3}$) is not caused by the valley-orbit interaction but by wavefunction overlap between the adjacent rows of phosphorus atoms. The $\Gamma_{1}-\Gamma_{3}$ splitting is then used to calculate the electronic extent of the single-row wire, which is found to be at least 4.6~nm. When two single-row wires are separated by in-plane distances less than 4.6~nm, the resulting energy splittings reduce the number of available conducting modes and, therefore, conductance at low bias voltages.

Furthermore, we present a nonequilibrium Green's function model for electron transport in a Si:P $\delta$-doped wire. We calculate the I-V characteristics of a variety of $\delta$-doped wires which have different in-plane widths, achieving good agreement with experiment. These wires show a linear response to an applied bias voltage, which is characteristic of the metallic conduction that is observed in experiment. We also show that the conductance can be controlled by applying a gate voltage to the wires. In the future this NEGF model could be extended to include donor disorder and non-ballistic transport.

\acknowledgements{The authors acknowledge the support of the NCI National Facility (Canberra, Australia).}

\appendix\section{\label{sec:appendix}Band folding in $\delta$-doped wires}

We analyze the band structure of the $\delta$-doped wires with respect to the band structure of bulk silicon. It is well-known that silicon is an indirect bandgap semiconductor with a sixfold degenerate CB edge~\cite{Madelung1996a}. These six CB edges are located at $k_{0}\approx0.85\frac{2\pi}{a}$ in the first BZ of the face-centered cubic (FCC) Bravais lattice, one along each of the six $\bracket{100}$ directions (where $a$ is the lattice constant of bulk silicon). The band structure of bulk silicon calculated using a 2-atom FCC unit cell is shown in Fig.~\ref{fig:bulk-bands}, where $X_{\mathrm{FCC}}$ is a point of high symmetry in the first BZ of the FCC unit cell and the path $\Gamma \to X_{\mathrm{FCC}}$ lies along one of the $\bracket{100}$ directions in reciprocal space~\cite{Bradley1972a}. The CB edges are located at $k_{0}\approx0.85X_{\mathrm{FCC}}$ in Fig.~\ref{fig:bulk-bands} as $|X_{\mathrm{FCC}}|=\frac{2\pi}{a}$. The band dispersion in the neighborhood of the CB edges is approximately parabolic~\cite{Drumm2012a,Drumm2013a} and, therefore, in reciprocal space these CB edges can be represented in 3D by spheroidal surfaces of constant energy centered at $k_{0}$~\cite{Singleton2001a} as shown in Fig.~\ref{fig:valleys}{a}. The spheroidal surfaces are anisotropic because the curvature of the band dispersion in silicon is anisotropic.

\begin{figure}[t!]
    \centering
    \includegraphics[]{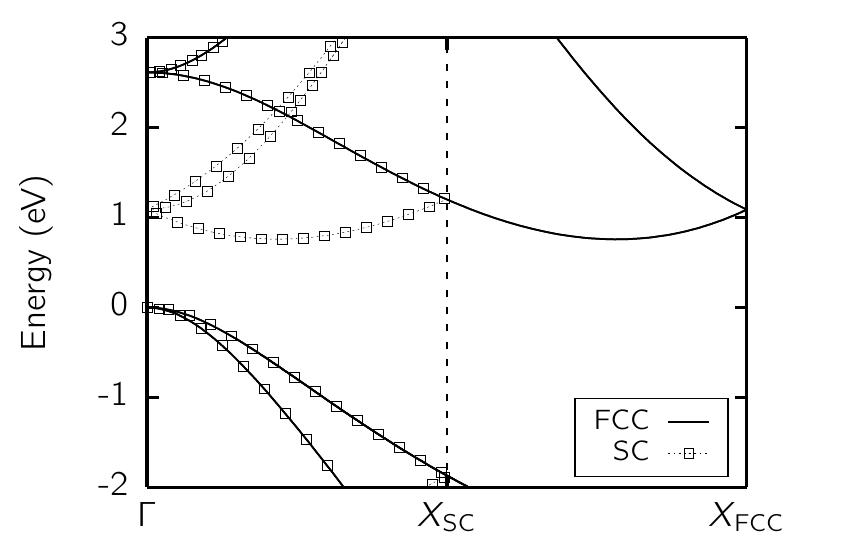}
    \label{fig:bulk-bands}
    \caption{The band structure of bulk silicon calculated using a 2-atom face-centered cubic (FCC) unit cell (solid lines) and an 8-atom simple cubic (SC) unit cell (squares). These band structures were calculated using the method described in Section~\ref{sec:methods-dft} with a $6\times6\times6$ Monkhorst-Pack $k$-point grid.}
    \label{fig:bulk-bands}
\end{figure}

\begin{figure*}[t!]
    \centering
    \includegraphics[]{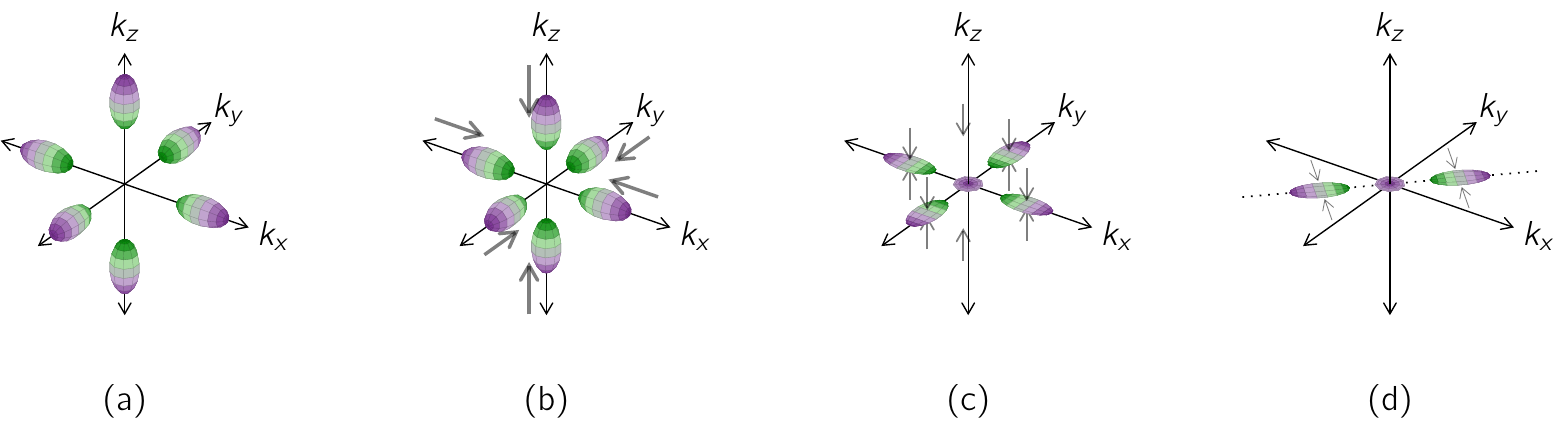}
    \caption{(color online) A 3D representation of the six-fold degenerate CB minimum of bulk silicon showing the CB valleys as spheroidal surfaces of constant energy centered at $k_{0}$ for (a) a FCC unit cell and (b) the same representation for a SC unit cell, where the spheroids have been translated along each of the cardinal $k$ axes towards $k=\left(0,0,0\right)$. (c) A 2D representation of the six CB minima for a TET supercell, where the spheroids in (b) have been folded to the $k_{x}k_{y}$ plane and (d) a 1D representation of the six CB minima for an ORT supercell, where the ellipses in (c) have been folded onto the line $k_{x}=k_{y}$. The width of these ellipses perpendicular to the line $k_{x}=k_{y}$ go to zero in the limit as the length of the ORT supercell in the $\left[1\bar{1}0\right]$ direction tends to infinity. These surfaces have been colored as a guide to the eye only, there is no other meaning intended by the colors.}
    \label{fig:valleys}
\end{figure*}

Fig.~\ref{fig:bulk-bands} also shows the band structure of bulk silicon calculated using an 8-atom simple cubic (SC) unit cell. From a comparison of the two band structures, we see the location of the CB edges and, therefore, spheroids in reciprocal space is dependent on the real-space unit cell that is used for the calculation, which is a result of band folding as discussed in our earlier work (see Appendices 1 and 2 of Ref.~\onlinecite{Drumm2013b}). In Fig.~\ref{fig:bulk-bands}, the BZ is folded about $k=\frac{\pi}{a}$ due to a doubling in the length of the supercell in the $\left[100\right]$ direction from 2.73~\AA~(for the FCC unit cell) to 5.46~\AA~(for the SC unit cell). When the length of the unit cell is increased in one dimension, the length of the BZ in the equivalent reciprocal dimension is decreased. The CB edges are thereby folded along their corresponding reciprocal space dimension towards the $\Gamma$ point\footnote{In general, when length of the BZ is decreased in one dimension, the distance between the CB edge and the $\Gamma$ point is not certain to decrease (this relationship is not monotonic). Rather, this distance decreases on average as the length of the BZ in this dimension is decreased.} at $k=\left(0,0,0\right)$. The CB edges have been folded from ${k_{0}\approx0.85\frac{2\pi}{a}}$ to ${k_{0}\approx0.15\frac{2\pi}{a}}$ in Fig.~\ref{fig:bulk-bands} and this is also shown in Fig.~\ref{fig:valleys}{b} by the translation of the spheroids along each of the cardinal $k$ axes towards $k=\left(0,0,0\right)$.

The band structure of bulk silicon calculated using a 1280-atom orthorhombic (ORT) supercell (for example, see Fig.~\ref{fig:1-row}{b}) is shown as the gray shaded region in Fig.~\ref{fig:1-row-bands}. There are two CB edges at energy zero; one at $\Gamma$ and the other at $k\approx0.16\frac{2\pi}{a}$. These are each doubly degenerate and, therefore, represent four of the six CB edges of bulk silicon. The other two CB edges are located at $k\approx-0.16\frac{2\pi}{a}$ and are not shown as they are symmetrically equivalent to those at $k\approx0.16\frac{2\pi}{a}$. The locations of the CB edges in reciprocal space are dependent on the supercell that is used for the calculation and the resulting band folding of the SC (and ultimately FCC) band structure. Therefore, the location of the CB edges for the 1280-atom ORT supercell can be predicted from the band structure calculated using the 8-atom SC unit cell and simple geometric arguments.

To see how the SC band structure can be used to calculate the location of the CB edges for the 1280-atom ORT supercell, consider the simulation cell for an Si:P $\delta$-doped layer. We use a 16-atom tetragonal (TET) unit cell to represent $\delta$-doped layers because the phosphorus atoms are doped in-plane at densities of 0.25~ML and the supercell needs to include at least four silicon atoms in the donor plane (so one of these atoms can be substituted by a phosphorus atom for a doping density of one in four)~\cite{Budi2012a,Drumm2013b}.

The TET unit cell is rotated by $45^{\circ}$ about the $\left[001\right]$ axis compared to the 8-atom SC unit cell. This rotation does not affect the location of the CB edges in reciprocal space, only their relative position in the first BZ of the TET unit cell. The length of the TET unit cell in the $z$ (\textit{i.e.} $\left[001\right]$) direction is greater than that of the SC unit cell and this folds the CB edges in the $k_{z}$ direction towards $\Gamma$. If the length of the TET unit cell in the $\left[001\right]$ direction is increased, it becomes a TET supercell. If this TET supercell is large enough to separate the $\delta$-doped layer from its periodic images in the $\left[001\right]$ direction, then the CB edges in the $k_{z}$ direction are folded to the $\Gamma$ point\footnote{The CB edges are only approximately folded to $\Gamma$. This approximation is only exact in the limit as the length of the TET unit cell in the $\left[001\right]$ direction tends to infinity.}. This folding is shown as a ``flattening'' of the spheroids to the $k_{x}k_{y}$ plane in Fig.~\ref{fig:valleys}{c}. In this figure, there are CB edges located at $\Gamma$ and ${k\approx0.15\frac{2\pi}{a}}$ in the $\left[100\right]$ direction and, indeed, these have previously been reported for a TET supercell~\cite{Drumm2013b}.

The length of the simulation cell must also be large in the $\left[1\bar{1}0\right]$ direction for an Si:P $\delta$-doped wire so the 1D confinement of the donor electrons can be modeled accurately. In addition, the length of the simulation cell in the $\left[1\bar{1}0\right]$ direction will be different to the length of the simulation cell in the $\left[001\right]$ direction (for atomistic simulations) because of the different crystallographic symmetries in each of these directions. Therefore, we must use an ORT supercell rather than a TET supercell to simulate the $\delta$-doped wires.

The ORT supercell is elongated in the $\left[1\bar{1}0\right]$ direction compared to the TET supercell, which shortens the equivalent reciprocal space dimension of the first BZ (parallel to the $\left[1\bar{1}0\right]$ $k$-space direction and perpendicular to the line $k_{x}=k_{y}$). The four CB edges at $k\approx0.15\frac{2\pi}{a}$ in Fig.~\ref{fig:valleys}{c} are thereby folded along the $\left[1\bar{1}0\right]$ $k$-space direction towards the line $k_{x}=k_{y}$, as is shown in Fig.~\ref{fig:valleys}{d}. Two of the six CB edges are located at $k\approx\frac{1}{\sqrt{2}}0.15\frac{2\pi}{a}\approx0.11\frac{2\pi}{a}$ along the line $k_{x}=k_{y}$ in Fig.~\ref{fig:valleys}{d} and another two at $k\approx-0.11\frac{2\pi}{a}$ along the same line. There is also a $\frac{1}{\sqrt{2}}$ contraction of the valleys for these four CB edges as a result of this folding and, ultimately, the $45^{\circ}$ rotation of the ORT supercell (relative to the SC unit cell)~\cite{Drumm2013a}. This $\frac{1}{\sqrt{2}}$ contraction of the valleys causes an effective doubling in the curvature of the band dispersion and, therefore, a reduction of $\frac{1}{2}$ in the corresponding effective mass~\cite{Drumm2013a}. In addition, the CB edges are only approximately folded onto the line $k_{x}=k_{y}$. This approximation is only exact in the limit as the length of the ORT supercell in the $\left[1\bar{1}0\right]$ direction tends to infinity. Therefore, the CB edge along the path $\Gamma\to X_{\mathrm{ORT}}$ in Figure~\ref{fig:1-row-bands} is located at $k\approx0.46X_{\mathrm{ORT}}\approx0.16\frac{2\pi}{a}$ rather than $k\approx0.11\frac{2\pi}{a}$ (where $X_{\mathrm{ORT}}=\frac{1}{2\sqrt{2}}\frac{2\pi}{a}$ in the $\left[110\right]$ $k$-space direction). For double-row wires C and F, where the length of the ORT supercell in the $\left[1\bar{1}0\right]$ direction is larger, the same CB edge is located at $k\approx0.38X_{\mathrm{ORT}}\approx0.13\frac{2\pi}{a}$ in the $\left[110\right]$ $k$-space direction\footnote{For double-row wires C and F, the ORT supercell is elongated in the $\left[1\bar{1}0\right]$ direction so that the amount of silicon cladding perpendicular to the phosphorus wires is not less than 2.7~nm.}.

\bibliography{NWp2.bib}

\end{document}